# Solid-to-solid phase transition from amorphous carbon to graphite nanocrystal induced by intense femtosecond x-ray pulses.

J. Gaudin,[1,*] J. Chalupský,[2,3] M. Toufarová,[2,3] L. Vyšín,[2] V. Hájková,[2] R. Sobierajski,[4,5] T. Burian,[2] Sh. Dastjani-Farahani,[1] A. Graf,[6] M. Amati,[7] L. Gregoratti,[7] S.P. Hau-Riege,[6] G. Hoffmann,[8] L. Juha,[2] J. Krzywinski,[9] R. A. London,[6] S. Moeller,[9] H. Sinn,[1] S. Schorb,[9] M. Störmer,[10] Th. Tschentscher,[1] V. Vorlíček,[2] H. Vu,[8] J. Bozek,[9] and C. Bostedt[9]

[1]European XFEL GmbH, Albert-Einstein-Ring 19, D-22761 Hamburg, Germany

[2]Institute of Physics, Academy of Sciences of the Czech Republic, Na Slovance 2, 182 21 Prague 8, Czech Republic

[3]Faculty of Nuclear Sciences and Physical Engineering, Czech Technical University in Prague, Břehová 7, 115 19 Prague 1, Czech Republic

[4]Institute of Physics PAS, Al. Lotników 32/46, PL-02-668 Warsaw, Poland

[5]FOM -Institute for Plasma Physics Rijnhuizen, NL-3430 BE Nieuwegein, The Netherlands

[6]Lawrence Livermore National Laboratory, Livermore, California 94550, USA

[7]Sincrotrone Trieste SCpA, SS14-Jm Area Science Park, 34149 Trieste, Italy

[8]Institute of Applied Physics, University of Hamburg, D-20355 Hamburg, Germany

[9]SLAC National Accelerator Laboratory, 2575 Sand Hill Road, Menlo Park, California 94025, USA

[10] Helmholtz Zentrum Geesthacht, Institute of Materials Research, Max-Planck-Str. 1, D-21502 Geesthacht, Germany

Corresponding author. Tel/fax: +49 8998 5456/1905.  E-mail address: jerome.gaudin@xfel.eu (J. Gaudin)

Abstract:

We present the results of an experiment where amorphous carbon was irradiated by femtosecond x-ray free electron laser pulses. The 830 eV laser pulses induce a phase transition in the material which is characterized ex-situ. The phase transition energy threshold is determined by measuring the surface of each irradiated area using an optical Nomarski microscope. The threshold fluence is found to be 282 ± 11 mJ/cm², corresponding to an absorbed dose at the surface of 131 ± 5 meV/atom. Atomic force microscopy measurements show volume expansion of the irradiated sample area, suggesting a solid to solid phase transition. Deeper insight into the phase transition is gained by using scanning photoelectron microscopy and micro-Raman spectroscopy. Photoelectron microscopy shows graphitization, i.e. modification from $sp^3$ to $sp^2$ hybridization, of the irradiated material. The micro-Raman spectra show the appearance of local order, i.e. formation of graphite nanocrystals. Finally, the nature of the phase transition is discussed, taking into account previous theory and experimental results.

## 1. Introduction

Carbon has the very interesting property of forming different structures, named allotropes, due to the different hybridizations $sp^3$, $sp^2$ and sp. The synthesis and the characterization of these allotropes have opened a wide range of applications leading to the advent of a new Carbon era [1]. Amorphous carbon (a-C) is one of these synthetic allotropes, made of a mixture of tetrahedral $sp^3$ (diamond-like) and trigonal $sp^2$ (graphite-like) carbon in various ratios. For high $sp^3$ content, this material is called diamond-like carbon. Its physical properties (such as mechanical and radiation hardness, biocompatibility, and chemical inertness…) are useful in numerous domains ranging from x-ray optics to micro-electronics (see ref. [2] for a comprehensive review). Phase transitions in a-C have also been investigated using different types of triggering processes: thermal treatment [3,4,5] or irradiation by ions [6], electrons [7] and optical laser pulses [8]. All studies demonstrated that the a-C structure tends to be modified to a more ordered graphitic phase, e.g. by undergoing an amorphous → crystalline transition.

The advent of x-ray free electron lasers (XFEL), also known as 4$^{th}$ generation x-ray light source, now enables the study of such phase transitions induced by ultra-short, femtosecond, intense x-ray pulses. The advantage of x-ray radiation is the large penetration depth, in the micrometre range, implying that a large volume of material can be submitted to radiation, contrary to optical wavelengths. Moreover the effect of core level ionization can also be studied.

We present the results of an experiment conducted at the Linac Coherent Light Source (LCLS) [9] where a-C was irradiated by a 830 eV x-ray FEL beam. In the first part of this article, we describe the experiment and the procedure used to determine the phase transition energy threshold by using ex-situ optical microscopy. The second part focuses on scanning photoelectron microscopy (SPEM) and micro-Raman spectroscopy analysis. These two analysis tools are complementary as SPEM provides information on the chemical bonding while Raman spectroscopy unravels short range atomic order. Finally, the question of whether material modification is of thermal or non-thermal nature is discussed.

2. **Experiment and determination of phase transition energy threshold**

The experiment was performed at the atomic, molecular and optical science instrument at the LCLS. The 830 eV photon beam was focused with a Kirkpatrick Baez mirrors system onto the sample at normal incidence angle. The irradiation was performed under vacuum. After each laser pulse, the sample was moved to a fresh spot. The single pulse energy was measured up to 30 µJ on the sample, and was varied using a gas attenuator. The pulse duration was in the sub 100 fs range. The sample consisted of a 1.4 µm thick a-C coating on a Si substrate. The a-C was deposited by magnetron sputtering and x-ray reflectivity measurements indicated a density of 2.2 g/cm³ [10] corresponding to a ratio of $sp^3 / sp^2$ of 0.2 [2].

In order to determine the phase transition threshold, we used optical Nomarski microscopy, which is sensitive to variation of optical refractive index, hence to any phase change, evidenced by a change of color in the microscope measurement. For each irradiated sample region, the area showing a color change on the microscope image was measured. Figure 1 shows the dependence of the area versus incident energy near the transition threshold.

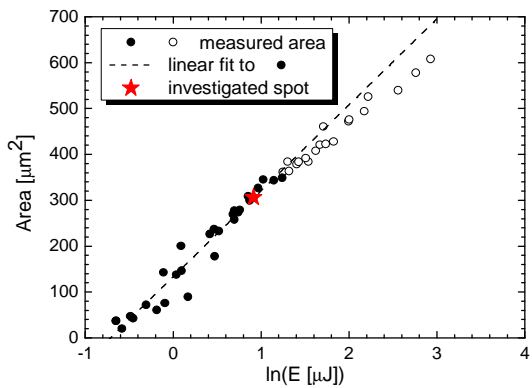

Figure 1: Surface area of the irradiated materials as a function of impinging energy as measured with Nomarski microscope. The red star corresponds to the spot shown in Fig. 2

The energy threshold is found to be 0.49 ± 0.2 µJ by fitting the experimental points with a logarithmic function as described in ref [11]. Due to the non-Gaussian shape of the beam, only the very first points (the filled-black ones in Figure 1) are used for the fitting procedure. In order to determine the corresponding fluence threshold the beam area is needed. Since the spatial profile of the beam is non-Gaussian, a specific method described in details in ref. [12] has been applied to retrieve the effective area. This surface is equivalent to the area at 1/e level of a Gaussian beam. The effective area is then equal to 174 µm², leading to a fluence threshold equal to 282 ± 11 mJ / cm². Another useful quantity is the averaged surface dose absorbed per atom. Taking into account the photoabsorption cross section of 3706 cm²/g[1], the surface dose is found to be 131 ± 5 meV / atom. In comparison, this threshold is lower than for bulk $B_4C$ where the threshold was found to be 684 meV / atom for similar irradiation conditions [13].

Further insight is gained with atomic force microscopy (AFM) measurements. Figure 2 shows an example for a shot with incident energy of 2.5 µJ (red star in Figure 1). It appears that the irradiated material undergoes a volume expansion with no ablation. The AFM measurements also provide an effective way to estimate the local fluence value. In fact, assuming that there is no transverse energy transfer, we can assign the iso-height contours and their respective areas a fluence value, knowing the total pulse energy and the so called fluence scan of the beam profile [12]. Therefore one height value corresponds to one local fluence value within a confidence interval of 20%. The resulting fluence map is shown in Figure 2 and will be used in the later sections.

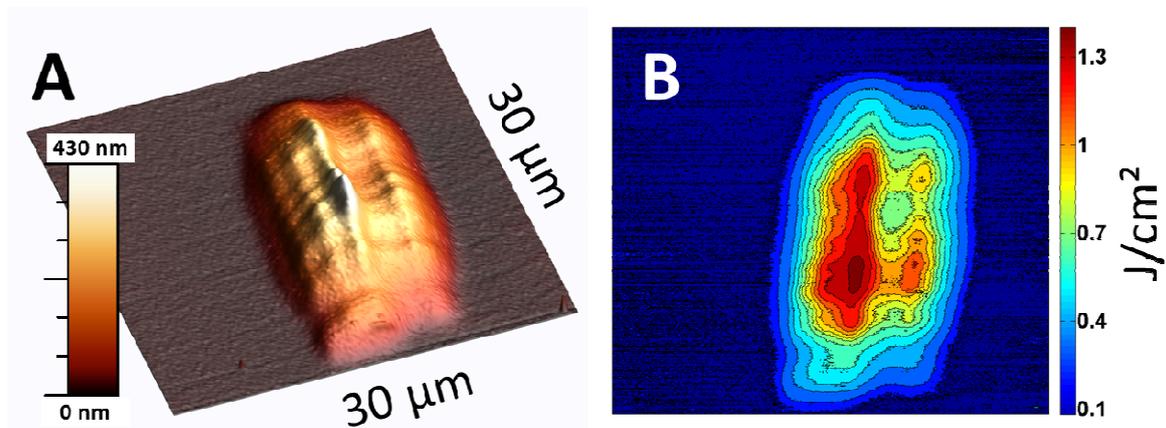

**Figure 2 AFM topography of an irradiated area with a 2.5 µJ FEL pulse. B: Fluence map deduced from the AFM measurement. The color scale is in J/cm²**

## 3. Chemical and structural analysis

---

[1] Center for X-Ray Optics - http://henke.lbl.gov/optical_constants/.

The irradiation by an XFEL pulse at moderate fluence triggers a solid-to-solid phase transition. In order to gain more insight in the nature of the modification, two types of analysis were performed ex-situ: SPEM and micro-Raman analysis. SPEM provides information on the chemical bonding while Raman spectroscopy unravels short range atomic order.

*3.1. Scanning Photo Electron Microscopy*

Photoemission spectroscopy (PES) is a well-established technique to characterize a-C [14, 15], and its $sp^3$ and $sp^2$ content. SPEM allows the same type of analysis in a space-resolved way, as it is an imaging technique. The experiment was performed at the ESCA microscopy beamline at the Elettra synchrotron facility. The complete description of the beamline is given in ref. [16]. Briefly, a 649 eV x-ray beam is monochromatized and then focused on the sample with a zone plate optic, producing a circular spot with a diameter of about 150 nm. The sample is then raster scanned with respect to the focused beam. The photoemitted electrons are then collected by a hemispherical sector spectrometer and measured by a multichannel detector. For each pixel, the acquisition is done in a spectro-imaging mode, whereby the energy-dispersed electrons are measured by a 48-channel detector. As a result, photoemission spectrum over a limited energy range is also measured for each point of the image. Figure 3 shows the C 1s core level photoemission spectrum resulting from an average over 10 x 10 pixels for two different regions corresponding to a non-irradiated area and the area shown in Figure 2, which is irradiated at a fluence of 1.4 J/cm². The content of sp² and sp³ hybrids can be deduced by fitting the peak with two components: one corresponding to $sp^2$ hybrids around 284 eV, and a second one corresponding to $sp^3$ hybrids at 0.9 eV higher binding energy [14,15].

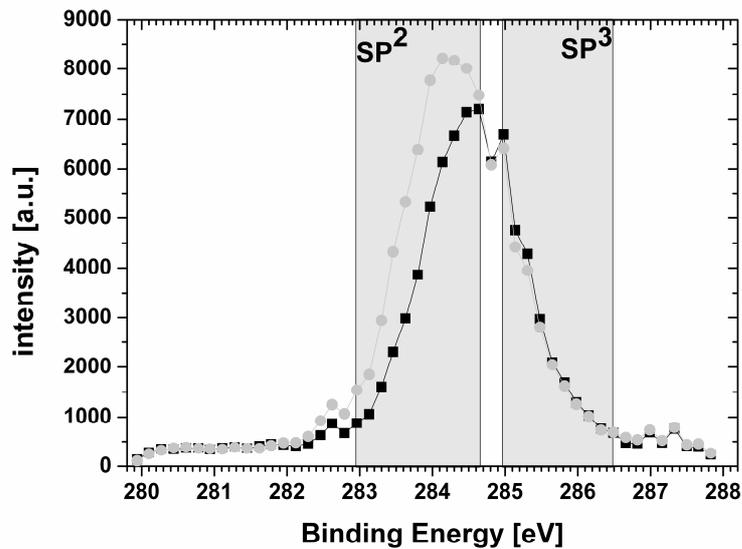

Figure 3: Photoemission spectra from SPEM measurements obtained from averaging over two 10x10 pixel areas. Black curve: non-irradiated sample – Grey Curve: center of the irradiated sample corresponding to 1.4 J/cm²

The irradiation induces a shift of the spectrum towards lower binding energy and a slight amplitude decrease of the high binding energy part, similar to what has been observed from an annealed a-C sample [14].

We define two energy ranges: the first one [282.9 eV to 284.7 eV] where the signal is mainly due to $sp^2$ hybrids and the second one [284.9 eV to 286.7 eV] corresponding to $sp^3$ hybrids. The chemical information is extracted from the SPEM images in the following way: the images resulting from the integration of the signal over these particular ranges are divided by the image resulting from the sum of all the channels. The resulting images are shown in Figure 4. The left figure corresponds to $sp^2$ and the right one to $sp^3$ content. The $sp^2$ component is higher in the irradiated area, whereas the $sp^3$ is lower indicating the conversion of the $sp^3$ to the $sp^2$ bonds, e.g. graphitization of the sample. Since the two components of the C1s peak overlap, the SPEM images give only a qualitative indication and not an absolute measurement of the hybrid contents.

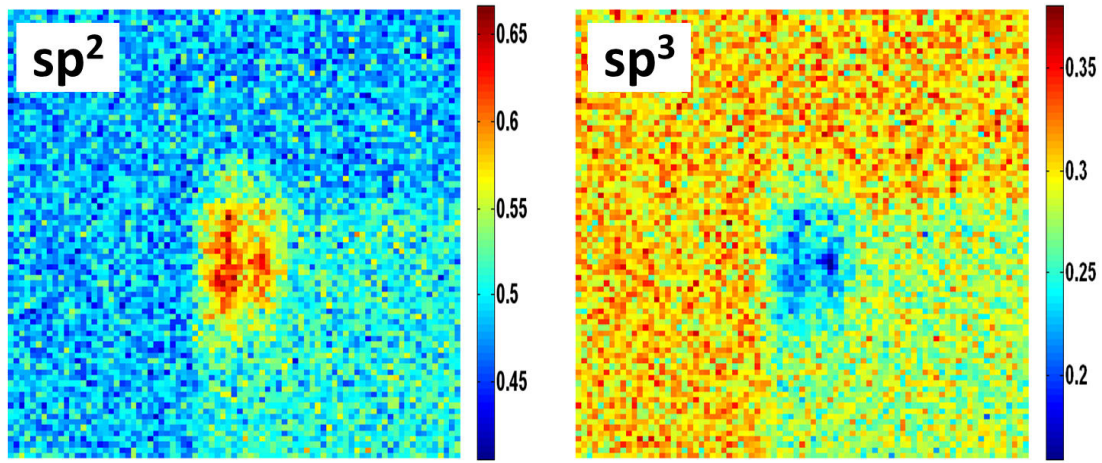

**Figure 4** <u>Left:</u> SPEM measurement corresponding to the fraction of sp² hybrids. <u>Right</u> SPEM measurement corresponding to the fraction of sp³ hybrids. The square in the right bottom part of the picture is the residual due to a previous scan measurement done to retrieve the exact location of the irradiated spot. The image dimensions are 89.6 x 89.6 μm.

*3.2. Micro-Raman analysis*

The micro-Raman analyses were performed prior to the SPEM measurements, in order to avoid any influence of the x-ray irradiation. The Raman spectra were measured in the usual backscattering geometry with a micro-Raman spectrometer (Renishaw Ramascope, Model 1000) coupled to an optical microscope focusing the $\lambda$ = 514.5 nm laser beam to a 2 μm spot diameter. At this wavelength only the sp² bonds are probed, due to their 50 to 250 higher scattering cross section as compared to sp³ [17] . Trial measurements were performed on test samples in order to determine the appropriate experimental conditions providing clean spectra without material modifications by laser annealing. Selected Raman spectra, corresponding to non-irradiated areas and areas irradiated at different fluences indicated in Figure 2, are shown in Figure 5. The spectra clearly show a broad asymmetrical feature typical of the a-C structure for the non-irradiated part. For the irradiated sample the two components of the initial feature clearly show up as the fluence increases. The first peak around 1350 cm$^{-1}$ (D peak) is related to defect and breathing modes of sp² rings while the second peak around 1600 cm$^{-1}$ (G peak) originates in the graphite related doubly degenerate phonon mode ($E_{2g}$ symmetry). The measured spectra are fitted using a Lorentzian line for the D peak and a Breit-Wigner-Fano (BWF) line for the G peak [18]. The BWF line shape has the advantage of being equivalent to a Lorentzian shape when $1/Q \rightarrow 0$, where Q is the BWF coupling coefficient. The spectrum corresponding to initial and irradiated material can then be fitted with the same line shape as shown in Figure 5.

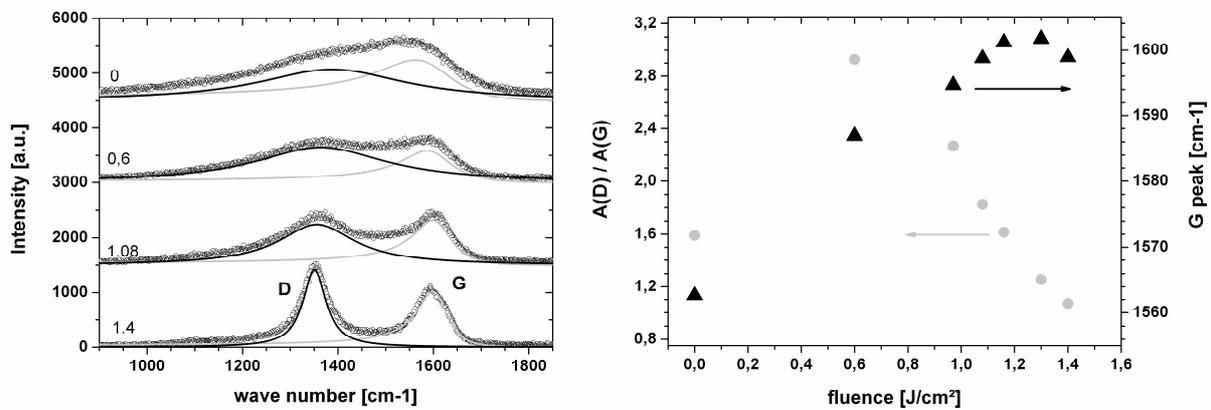

Figure 5 Left: Selected Raman spectrum for different fluence (from top to bottom: 0, 0.6, 1.08 and 1.4 J/cm²). The G (grey curve) and D (black curve) peaks used for the fitting are also shown. Right: Grey point: Area ratio of G and D peak resulting from the fit. Black triangles: Position of the G peak.

Two quantities are extracted from the fitting curves, as shown in Figure 5: the G peak position and the integrated peak intensity ratio A(D)/A(G). The right graph in Figure 5 shows the displacement of the G peak maximum from 1560 cm$^{-1}$ for the non-irradiated material up to 1600 cm$^{-1}$ at the maximum fluence, e.g. 1.4 J / cm². This can be explained by the decrease of the sp$^2$ bond length while the material turns to ordered graphite. This induces a bond strengthening and a phonon mode hardening reflected by a shift of the G peak [17]. The integrated peak intensity has been shown to be directly proportional to the in-plane size, denoted $L_a$, of graphitic nanocrystals. It can be considered as a measurement of the degree of order in the a-C sample. Two relations are possible, depending on the structure of the material

[18] : for a-C (small ordered regions) A(D)/A(G) ~ $L_a^2$, while for graphitic nano- and microcrystals A(D)/A(G) ~ $1/L_a$. The results shown in Figure 5 can then be explained as follows: The A(D)/A(G) ratio increases as the irradiation is increased up to 0.6 J/cm² as expected for the a-C structure. For higher fluences, the value of A(D)/A(G) decreases indicating that the material is mainly composed of growing graphitic nanocrystals. Considering the relation given in ref. [4] , for 1.4 J / cm² $L_a$, is approximatly 15 nm. This shows that the size of the graphite nanocrystals is growing while the fluence is increasing.

## 4. Discussion

The SPEM and Raman measurments allow full characterization of the phase transition induced by the x-ray FEL pulse. The SPEM spectra clearly show a $sp^3 \rightarrow sp^2$ transition, while the Raman spectra clearly show a structure changes corresponding to an amorphous → crystalline transition. Unraveling the path that leads from the initial to the final states is of course difficult. Previous experimental results and recent theory can provide qualitative arguments. Covalent materials, like carbon, have been shown to undergo ultra-fast non-thermal phase transition while irradiated by fs optical pulses [19, 20]. The basic picture is that the high density electron-hole plasma excited by the laser pulse tends to modify the potential energy surface, initiating atomic displacement. Providing that the original material has a local structure somewhat similar to the crystalline one, an amorphous to crystalline transition can then be achieved. In our specific case, as indicated by the high initial $sp^2$ content, the amorphous structure is mainly made of rings made of three to twelve carbon atoms [21] , randomly assembled. Recent theory demonstrates that the driving parameter in this type of transition is the electron-hole density [22]. In the case of optical fs pulses the energy is deposited in a thin layer (on the order of 100 nm) leading to an extremely high electron-hole density. In the specific case of x-ray radiation, the energy is absorbed in a much larger volume due to the micrometre range absorption depth. Nevertheless x-ray photons create photo- and

Auger electrons with high kinetic energy, which will relax mainly by ionizing secondary electrons. We estimate the ionization level of the sample in the following way: the incident flux of photons at the phase transition threshold is estimated to $F_{ph} = 2.14\ 10^{15}\ ph/cm^2$. In carbon based materials, e.g. diamond, it has been shown that a single primary electron, with a kinetic energy of 546 eV corresponding to a 1s electron ionized by a 830 eV photon, lead to the excitation of around $N_{2^{ndary}} \approx 70$ secondary electrons [23]. The density of free electron is then estimated to be $N_{e^-} = F_{ph} \cdot N_{2^{ndary}} / d_a = 1.25\ 10^{21}\ e^-/cm^3$ which is nearly 1 % of the atomic density is. This ionization level is of the order of magnitude required for non-thermal phase transition.

On the other hand, thermal treatment can induce the same kind of phase transition we observed, for a-C with low sp$^3$ content [5] as well as for high sp$^3$ content [3, 4] . The question arises then, if the x-ray irradiation allows reaching a similar temperature? The conversion from dose to temperature can be estimated by assuming that the dose is equal to the change in enthalpy in going from room temperature ($T_0$) to the elevated temperature ($T_1$): $\Delta H = \int_{T_0}^{T_1} C_p(T) \cdot dT$ . In this case, the assumption is made that all the absorbed x-ray energy is converted into heat. For the calculation, we used the value of the heat capacity $C_p(T)$ of graphite from reference [24] . The results are presented in Figure 6.

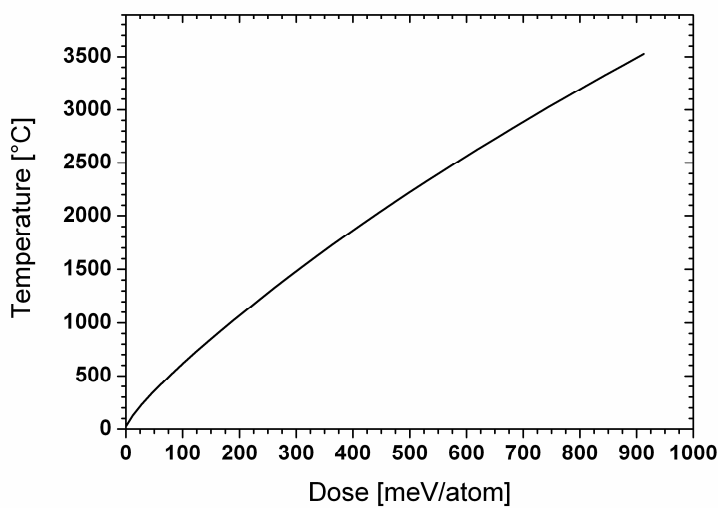

Figure 6 Temperature vs dose obtained from the enthalpy calculation

From this calculation, the temperature corresponding to the dose threshold (132 meV/atom) is found to be nearly 800°C. For a-C with an initial $sp^3$ content of 10%, close to the content of our sample, the formation of graphitic plane has been observed for T > 600°C [5]. So the heating due to the x-ray pulse is in theory high enough to trigger the phase transition.

## 5. Conclusion

We have measured the absolute dose for phase transition in a-C induced by femtosecond XFEL pulses to be 132 ± 5 meV / atom. Several types of analyses performed on the irradiated sample show the phase transition from amorphous to crystalline. At low fluence the initial $sp^2$ bonded atoms start to form graphite nanocrystals. The $sp^3$ bonds can be considered as defect in this new graphite structure. At higher fluence value, the graphitization of the $sp^3$ bonds, evidenced by the SPEM measurement, allows the formation of larger nanocrystals. This hypothesis is supported by the micro-Raman measurements which show that the size of the nanocrystals becomes larger as the incident fluence is increased. The maximum size of the graphite we measured is 15 nm.

The nature of the transition is still unclear. The qualitative arguments, based on previous experiments on covalent materials and recent models, indicate that the ionization by the intense x-ray pulse is high enough to trigger a non-thermal phase transition. On the other hand, the temperature estimated by assuming that all the x-ray energy is converted in heat, is high enough to thermally induce the local reordering we observed. At this stage, it is then difficult to draw a conclusion on the origin of the phase transition. Further theory development of the interaction of intense x-ray pulses with solids will be needed to gain deeper insight. On the experimental side, time-resolved experiments, which are now possible at x-ray FEL facilities, will also bring valuable information on the atomic dynamics of this type of transition.